\documentclass[12pt]{article}

\usepackage[a4paper, textwidth={17cm}]{geometry}

\usepackage{listings}

\usepackage{graphicx,float,wrapfig}

\usepackage{amsfonts}
\usepackage{amsmath}
\usepackage{amsthm}
\usepackage{amssymb}
\usepackage{authblk}

\usepackage[utf8]{inputenc}

\newcommand{\mket}[1]{| #1 \rangle}

\newcommand{\mbraket}[2]{\langle #1 | #2 \rangle}

\newtheorem{algorithm}{Algorithm}

\begin{document}

\lstset{language=C++}

\title{QTM: computational package using MPI protocol for quantum trajectories method}

\author[1]{Marek Sawerwain}
\author[2]{Joanna Wi\'sniewska}

\affil[1]{Institute of Control \& Computation Engineering University of Zielona G\'ora, Licealna 9, Zielona G\'ora 65-417, Poland; M.Sawerwain@issi.uz.zgora.pl}
\affil[2]{Institute of Information Systems, Faculty of Cybernetics, Military University of Technology, Urbanowicza 2, 00-908 Warsaw, Poland; jwisniewska@wat.edu.pl}

\date{}

\maketitle

\begin{abstract}
The Quantum Trajectories Method (QTM) is one of {the} frequently used methods for studying open quantum systems. { The main idea of this method is {the} evolution of wave functions which {describe the system (as functions of time). Then,} so-called quantum jumps are applied at {a} randomly selected point in time. {The} obtained system state is called as a trajectory. After averaging many single trajectories{,} we obtain the approximation of the behavior of {a} quantum system.} {This fact also allows} us to use parallel computation methods. In the article{,} we discuss the QTM package which is supported by the MPI technology. Using MPI allowed {utilizing} the parallel computing for calculating the trajectories and averaging them -- as the effect of these actions{,} the time {taken by} calculations is shorter. In spite of using the C++ programming language, the presented solution is easy to utilize and does not need any advanced programming techniques. At the same time{,} it offers a higher performance than other packages realizing the QTM. It is especially important in the case of harder computational tasks{,} and the use of MPI allows {improving the} performance of particular problems which can be solved in the field of open quantum systems.
\end{abstract}

\section{Introduction} \label{lbl:sec:introduction}

The Quantum Trajectories Method (QTM) is an important method actively applied for investigation in the field of open quantum systems \cite{Yip2018}, \cite{Liniov2017}, \cite{Daley2014}, \cite{Wyatt2005}. It was implemented as packages in a few programming languages and tools. The {first implementation} was a package written by Sze M. Tan for the Matlab environment \cite{SzeMTan1999}. There is also a package for the C++ language \cite{SchackaBrun1997}, \cite{SANDNER2014}. The latest implementations are QuTIP \cite{Johansson2012}, \cite{Johansson2013} for the Python Ecosystem and QuantumOptics.jl \cite{Kramer2018} prepared for the Julia language.

The {aforementioned} solutions, especially QuTIP and QuantumOptics.jl, allow utilizing parallel computing inside the environments of Python and Julia. However, the packages are not {intended} for High-Performance Computing (HPC) \cite{Pimenov2017}, \cite{Vasiliu2017} where Message Process Interface, termed as MPI \cite{Clarke1994}, plays a significant role. Using MPI allows {us} to re-implement the QTM for HPC systems, regardless of their scale (the QTM is scale-free because each trajectory may be calculated separately). The scale-free character of {the} QTM will allow {utilizing} more computing power, and that will result in shorter time of calculations, which is especially important for cases {in which} 20,000 and more trajectories are generated. 

Previously, we prepared the implementations of {the} QTM for CPUs and GPUs \cite{WisniewskaSawerwain2014}, \cite{Wisniewska2015}. The results presented in this work relate to a brand new implementation of {the} QTM for {the} MPI protocol ({the} actual source code of {the} QTM can be found at \cite{QTMRepo}). {The version 1.0a of the QTM package is also available at \cite{QTMZenodo}.}

{As} far as the QTM is concerned, a proper selection method for solving systems of Ordinary Differential Equations (ODEs) must be considered. This is a basic action taken while numerical computations are realized. More precisely, a very important issue is an Initial Value Problem (IVP). {Due to that, the} selection of {an} appropriate method for solving IVP (in general ODEs) is crucial. Especially when the system of equations is difficult to solve, e.g. so-called systems of stiff ODEs. The stiff ODEs constitute a significant type of ODEs and the correct solving of these equations is {pivotal} for numerical computations in many cases -- especially for {the} {QTM} where calculating {many trajectories requires solving many} ODEs.

A few groups of methods \cite{Hairer1993a}, \cite{Hairer1993b} used in the context of stiff ODEs may be recalled{,} and these are {the} Backward Differentiation Formula (BDF) methods.  Another approach, which is commonly used in numerical computations for solving ODEs, is Livermore Solver for ODE (LSODE) method  \cite{Hindmarsh:1980}. In this work{,} we reuse {the} LSODE variant called ZVODE (LSODE for complex numbers) for {the implementation of the QTM in the} MPI environment. 

The ZVODE package is a stable solution offering high accuracy, but this is not a reentrant solution which may be directly utilized in {a} parallel environment. However, this problem can be solved by utilizing MPI where the processes are {separate programs communicating with one another using message passing}. Therefore, one MPI process calls only one instance of {the} ZVODE method.   

The ZVODE package was implemented in the Fortran language. The QTM package, implemented in the C++ language, uses {the} ZVODE package efficiently. However, we made some effort to assure that functions offered by {the} QTM package are easily accessible as it is in {the} packages QuTIP and QuantumOptics.jl. Naturally, programs (delivered as examples or written by {the} user) utilizing {the} QTM have to be compiled. This should not be a problem because of used makefile mechanism for simplifying this process.

The paper {is organized} in the following way: in Section \ref{lbl:sec:QTM:ms:jw}{,} we shortly present selected mathematical features of {the} QTM. The BDF numerical methods for solving IVP, used in the presented package are discussed in {subsection} \ref{lbl:ssec:metods:for:IVPs:ms:jw}. Whereas in \ref{lbl:sec:qtm:algorithm:ms:jw} the algorithm for {the} QTM is presented. There are also some remarks pointing out where the parallel processing techniques and the MPI technology are utilized. 

Section~\ref{lbl:sec:impl:ms:jw} { contains selected remarks referring to the implementation of {the} QTM package. In this section{,} the most important data types implemented in package are presented. We also describe how the ZVODE package is used to solve ODEs {since} solving ODEs poses a significant problem in {the} QTM.}

We analyze the efficiency of our implementation in a comparison with {other recently developed packages} in section~\ref{lbl:sec:performance:results:PlosOne2018:ms:jw}. The most important issue {presented in this section is the scalability of the} QTM.

A summary of achieved results is discussed in Section~\ref{lbl:sec:conslusions:ms:jw}. There are also presented further aims which are planned to be realized as {the} next steps in {the} evolution of the demonstrated implementation of {the} QTM. The article is ended with acknowledgments and bibliography.

\section{Quantum Trajectories Method} \label{lbl:sec:QTM:ms:jw}

Performing the QTM requires solving ODEs. The presented solution, as mentioned above, uses {the} ZVODE package which allows {utilizing} some numerical methods indispensable for {the} proper functioning of {the} QTM. The fundamental information concerning {the} ZVODE package is presented in subsection~\ref{lbl:ssec:metods:for:IVPs:ms:jw}{, while} subsection~\ref{lbl:sec:qtm:algorithm:ms:jw} contains {the} description of {the} QTM, including {a method of calculating {a} single trajectory \cite{WisniewskaSawerwain2014}, \cite{Wisniewska2015} with {the} use of MPI} {  technology.} The properties of {the} QTM allow {us} to easily distribute the computations -- this feature helps to accelerate the calculations and makes the process scalable.  

Another important element of the solution is a {pseudorandom} number generator. In subsection \ref{lbl:ssec:prng:ms:jw}{,} we describe the generator which we chose to utilize in {the} implementation of {the} QTM package.

\subsection{Methods for solving ODEs} \label{lbl:ssec:metods:for:IVPs:ms:jw}

The ZVODE package was built on a basis of {the} VODE package  \cite{Brown1989}. The VODE package was based on {the} LSODE package. The numerical methods, used for the LSODE (Livermore Solver for Ordinary Differential Equations) package implementation, are the computational routines based on the group of Linear Multistep Methods (LMMs). In general, LSODE {uses} the Adams methods (so-called predictor-corrector methods) for the non-stiff ODEs solving. In case of stiff ODEs{,} the Backward Differentiation Formula (BDF) methods are used. In this section{,} the most important mathematical aspects of {the aforementioned} methods are {briefly} presented -- the detailed report concerning standard (i.e. sequential) implementation of LSODE may be found in \cite{LSODEA1}.

The choice of {the} ZVODE package is determined by the fact that {the} QTM needs using complex numbers and the ZDOVE package allows {performing} calculations on complex numbers of single and double precision.

For the Initial Value Problem (IVP) in general:
\begin{equation}
\begin{array}{lcl}
y' & = & f(x,y), \\
y(x_0) & = & y_0 .
\end{array}
\end{equation}
{The} LMMs, approximating the problem's solution, may be described as:
\begin{equation}
\sum_{i=0} ^{k} \alpha_{(k-i)} y_{(n+1-i)} = h \sum_{i=0}^{k} \beta_{(k-i)} f(x_{(n+1-i)}, y_{(n+1-i)}),
\end{equation}
where $\alpha_i, \beta_i \in \mathbb{R}$ and $\alpha_k \neq 0$ and $|\alpha_0| + |\beta_0| > 0$. The parameter $h$ represents the step width for integration. Naturally, a value of $h$ is selected during the solver's work with {the} use of adaptive methods.

The LSODE routine uses two methods based on {the} LMMs: {the} Adams-Moulton method and {the} BDF method.

The implicit Adams-Moulton method, this means the values of $\beta_i \neq 0$ (the detailed description of used symbols may be found in \cite{Soetaert2012}), may be presented as:
\begin{equation}
y_{n+1} = y_n + h \sum_{i=0}^{k} {\gamma_i} \triangledown^{i}  f_{n+1}
\end{equation}
where 
\begin{equation}
{\gamma_i} = {(-1)}^{i} \int_0^1  {-s +1 \choose i} ds
\end{equation}
The $\triangledown^{i}$ symbol stands for {the} backward differences. It should be pointed out that: $\triangledown^{i} f_n= f_n$ and $\triangledown^{i+1} f_n= \triangledown^{i}f_n - \triangledown^{i}f_{n-1}$. The value $k$ expresses the number of steps, i.e. the values of $y_i$ used in the specified method.

The BDF method is defined as:
\begin{equation}
\sum_{i=1}^{k} \frac{1}{i} \triangledown^{i } y_{n+1} = h f(x_{(n+1)}, y_{(n+1)})
\end{equation}
The above notation means that $\beta_k \neq 0$ but $\beta_i = 0$ for $i=0,1,2,\ldots,k-1$. The values of $\alpha_i$ are arbitrarily defined and may be found in many publications concerning the methods of solving ODEs.

In both cases, for {the} Adams-Moulton and {the} BDF method, the problem is to estimate the value of $f_{n+1}$. It should be stressed that this value is needed to estimate itself. The Newton method \cite{Press2007} is used in this case to calculate {the} estimation. The Newton method is quite rapidly convergent, though for the system of equations it needs to calculate the Jacobian. This method for (ZV/LS)ODE may be expressed as: 
\begin{equation}
g(y^{(1)}_{n+1}) = - \left(I - \frac{\partial f}{\partial f}\right) (y^{(2)}_{n+1} - y^{(1)}_{n+1}) 
\end{equation}
where $y^{(2)}_{n+1}$ and $y^{(1)}_{n+1}$ stand for the next approximations of $y_{n+1}$. The function $g( \cdot )$ represents the function approximating values of $y_{n+1}$.

\begin{figure}
\includegraphics[width=0.95\textwidth]{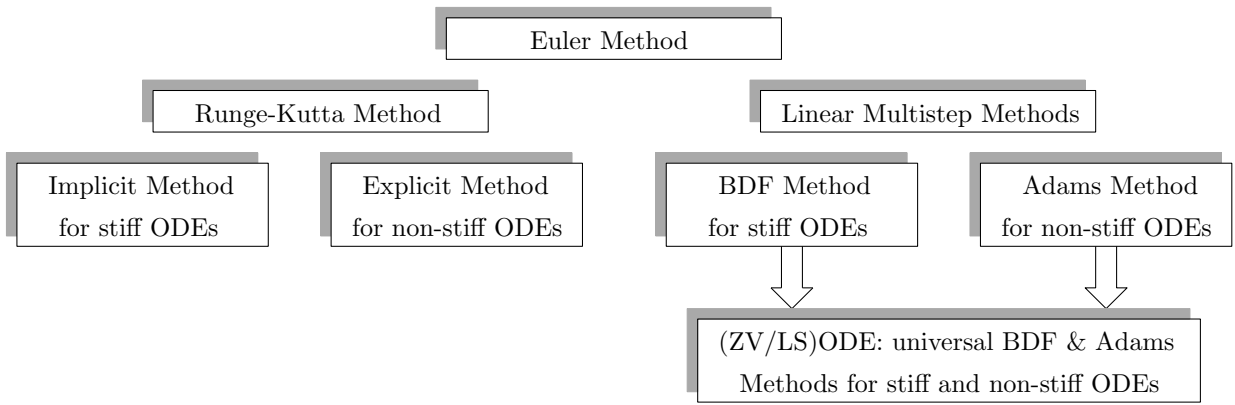}
\caption{{A general} overview of {the} ODE solver and {the} methods used in {the} (ZV/LS)ODE implementation}
\label{lbl:fig:lsode:metod:source}
\end{figure}

The BDF method is used for stiff ODE problems and the Adams-Moulton method for non-stiff ODE problems. Using these both approaches together makes the (ZV/LS)ODE a hybrid method -- see Fig.~\ref{lbl:fig:lsode:metod:source}. The presented QTM implementation allows {choosing} the method, respectively to solving easier (non-stiff ODE) or more difficult (stiff ODE) QTM {problems}. Moreover{,} in ZVODE, for every method the method's order {can also} be  controlled automatically. For {the} Adams-Moulton method{,} the orders from {the} first to {the} twelfth are available{,} and for {the} BDF method the orders from {the} first to {the} sixth.

\subsection{Quantum Trajectories Method} \label{lbl:sec:qtm:algorithm:ms:jw}

A description of quantum states' dynamics may be presented for two fundamental situations. The first case is an evolution of a closed quantum system{, whereas} the second case concerns an evolution of an open quantum system. As an example of a closed quantum system{,} we may refer to a model of {a} quantum circuit. If we want to consider a quantum system where its dynamics is affected by {the} influence of {an} external environment, we deal with an open quantum system. 

In this subsection{,} we do not aim to describe the mathematical models of dynamics in open and closed quantum systems -- {the details concerning this subject} may be found in \cite{Wyatt2005} and \cite{nielsen2010quantum}. {For clarity{,} it should be recalled} that for closed systems{,} the evolution is {a} unitary operation{,} and it can be denoted as Schr\"odinger equation:
\begin{equation}
(C1): \;\; i\hbar\frac{\partial}{\partial t}\Psi = \hat H \Psi, \;\;\; (C2): \;\; i\hbar\frac{d}{dt}\left|\psi\right> = H \left|\psi\right> ,
\end{equation} 
where (C1) is the form of {a} partial differential equation and (C2) is the form which is convenient to use in numerical simulations. In (C2) $H$ is a Hamiltonian {describing} system's dynamics{,} and $\mket{\psi}$ stands for the initial system's state. 
 
The von Neumann equation describes the quantum system's evolution if an influence of {an} external environment {has to be} considered:
\begin{equation}
\dot \rho_{\rm tot}(t) = -\frac{i}{\hbar}[H_{\rm tot}, \rho_{\rm tot}(t)], \;\;\; H_{tot}=H_{sys}+H_{env}+H_{int},
\label{eq:von:Neumann}
\end{equation} 
where $H_{sys}$ {denotes} the dynamics of {a} closed/{core} system, $H_{env}$ stands for the environment's dynamics{,} and $H_{int}$ {describes} the interaction's dynamics between the external environment and the system. { The environment's influence can be removed from (\ref{eq:von:Neumann}) by the partial trace operation. In such {a} case{,} we obtain {an} equation which describes {the} dynamics of the core system. Such {a} system can be expressed by the Lindblad master equation:}
\begin{equation}
\dot\rho(t)=-\frac{i}{\hbar}[H(t),\rho(t)]+\sum_n \frac{1}{2} \left[2 C_n \rho(t) C_n^{+} - \rho(t) C_n^{+} C_n - C_n^{+} C_n \rho(t)\right],
\end{equation} 
where $C_n$ {stands for} a set of collapse operators. These operators represent the influence of {an} external environment which affects a simulated system. Naturally, applying a collapse operator causes irreversible modification of a quantum state. It should be emphasized that simulating the behavior of {a} quantum system  {needs exponentially growing memory capacity according to the system's dimension.}

The QTM is a method which {facilitates reducing} the memory requirements during the simulation. Of course, any simulation of {a} quantum system's behavior needs calculating many single trajectories {-- the more, the better -- because they will {be} averaged to one final trajectory{,} and a greater number of trajectories ensures improved accuracy.} However, every trajectory may be simulated separately{,} and this fact {provides} a natural background to utilize a parallel approach while implementing {the} QTM.

If we would like to compare simulating the behavior of a quantum system with {the} use of Lindblad master equation and the QTM, we should consider the requirements of both methods on computational resources. The Lindblad master equation methods utilize the density matrix formalism and {the} QTM is based on a wave function of $n$-dimensional state's vector (termed as a pure state). A number of this vector's entries grows exponentially, but using sparse matrices {facilitates} efficient simulation of {a} quantum system's behavior. It should be stressed that a simulation based on the wave function concerns only one state of {a} quantum system what seems to be a disadvantage of this solution because for {the} Lindblad master equation methods{,} the density matrix describes many different states of the same system. Unfortunately, using density matrices in most cases is not possible because of memory requirements -- {the size of {the} density matrix grows exponentially when the dimension of {a} simulated system increases.} While the QTM { enables monitoring} the influence of {an} external environment on {a} quantum state by modifying the state's vector with {the} use of a collapse operator.
 
For {the QTM,} an evolution of {a} quantum system is described by so-called effective Hamiltonian $H_{\rm eff}${, defined with {the} use of a set of collapse operators $C_n$,} and {a} system's Hamiltonian $H_{\rm sys}$:
\begin{equation}
H_{\rm eff} = H_{\rm sys} - \frac{i\hbar}{2}\sum_{n}C^{+}_{n}C_{n}.
\label{eqn:effective:hamiltonian}
\end{equation}

The set of operators $C_n$ {determines} the probability of {a} quantum jump occurrence. This phenomenon is {caused by a single collapse operator acting on a current quantum state of the system. A probability of {a} quantum jump occurrence is:}
\begin{equation}
\delta p = \delta t \sum_{n}\left<\psi(t)|C^{\dagger}_{n}C_{n}|\psi(t)\right>.
\label{eqn:probabil:collapse}
\end{equation}

If a quantum jump takes place, the system's state {at} the moment of time $(t+\delta t)$ -- that is just after the collapse operation -- can be expressed as:
\begin{equation}
 \left|\psi(t+\delta t)\right>= \frac{C_{n}\left|\psi(t)\right>}{\sqrt{\left<\psi(t)|C_{n}^{\dagger}C_{n}|\psi(t)\right>}}.
 \label{eqn:state:for:new:time}
\end{equation}

Furthermore, if many collapse operators may be applied in {the} considered model, the probability of using $i$-th operator is:
\begin{equation}
 P_{i}(t)=\frac{\left<\psi(t)|C_{i}^{\dagger}C_{i}|\psi(t)\right>}{\delta p}.  
 \label{eqn:probability:of:cn}
\end{equation}

Of course, simulating the phenomenon of {a} system's collapse needs a random numbers generator to ensure its probabilistic character (this issue is discussed in the following subsection).

Let us emphasize that all {the} considered calculations correspond to operations implemented on matrices and vectors within the QTM package. {the needed} matrices are usually band matrices. {Namely, it is convenient, in terms of {the} memory consumption and {the} speed of calculation, to use sparse matrices. In consequence, the compressed sparse row (CSR) format may be utilized} -- it also gives an additional speed-up, especially when many matrix-vector multiplications have to be realized.

{By summarizing the above remarks,} we can formulate an algorithm which presents how a single quantum trajectory is calculated: 
 
\begin{algorithm} Computation of a single trajectory \\
The algorithm computing a single trajectory, according to {the} QTM, may be {described} as four computational steps:
\begin{enumerate}
\item[(A)] a value $ 0 \leq r < 1$ is computed by a pseudorandom number generator ($r$ denotes the probability of {a} quantum jump occurrence);
 
\item[(B)] to get the state's vector {at} the moment $t$ the Schr\"odinger equation is integrated with the Hamiltonian $H_{\rm eff}$ (providing that the state's vector norm {has} to be equal or greater to $r$: $\mbraket{\psi(t)}{\psi(t)} \geq r$);

\item[(C)] if a quantum jump is realized{,} then the system's state projection at the moment $t$, to one of the states given by Eq~(\ref{eqn:state:for:new:time}), is calculated. The operator $C_n$ is selected to meet the following relation for {the} adequate $n$: $\sum_{i=1}^{n} P_{i}(t) \ge r$ and $P_{i}(t)$ is given by Eq~(\ref{eqn:probability:of:cn});

\item[(D)] the state obtained by the projection of wave-function in the previous step is a new initial value corresponding to the moment of time $t$; next, the new value of $r$ is randomly selected{,} and the procedure repeats the process of {the} quantum trajectory generation{,} starting from the step (B) -- more precisely: the simulation is performed again{,} but starts from {the} previously given value of $t$. 
\end{enumerate}
\label{lbl:alg:for:single:qtm:PlosONE:ms:jw}
 \end{algorithm}
  
The presented approach is based on \cite{Johansson2012}, \cite{Johansson2013}, {\cite{WisniewskaSawerwain2014}, \cite{Wisniewska2015},} \cite{Garraway1994b}, \cite{Dum1992}.

\subsection{Pseudorandom number generator} \label{lbl:ssec:prng:ms:jw}

The QTM package described in this work utilizes a pseudorandom number generator from {the} Generalized Feedback Shift Register (GFSR) class. More precisely, we chose the LFSR113 method which is defined {by} using recurrence over the field $F_2$ consisting of elements ${0,1}$:
\begin{equation}
x_n = (a_1 x_{n-1} + a_2 x_{n-2} + \ldots + a_k x_{n-k} ) \mod 2,
\end{equation}
where $a_{j_1}$ are generator's parameters ($j_1=1,...,k$) and $x_{j_2}$ are generator's seeds ($j_2=n-1,...,n-k$). The generator's period is $r=2^k -1$ if and only if the characteristic polynomial of recurrence:
\begin{equation}
P(z) = z^k - a_1 z^{k-1} - \ldots - a_k,
\end{equation}
is primitive.

The generated values, for $n \geq 0$, may be expressed as: 
\begin{equation}
u_n = \sum^{L}_{i=1} x_{ns+i-1} 2^{-i} ,
\end{equation}
where $s$ denotes the step size{,} and $L$ stands for the number of bits in a generated word. If $(x_0,x_1,\ldots, x_{k-1}) \neq 0${,} and $s$ is coprime to $r${,} then we obtain a periodic sequence of $u_n$ (with a period denoted as $r$).

Of course, the quality of {a} generator is determined by the sequence $x_n$. The proper choice of seeds is described in \cite{LEcuyer1999} where {it} is also shown that four seeds ($k=4$) are sufficient to generate {high-quality} pseudorandom numbers. The {LFSR113 generator's} realization is very fast because of the bitwise operations usage -- this feature also does not collide with ensuring a sufficient period for generated numbers: $2^{113}$. However, it should be emphasized that the selection of generator's seeds is crucial -- there are four initial values{,} and they have to be integers greater than 1, 7, 15 and 127, respectively.

\section{General implementation remarks} \label{lbl:sec:impl:ms:jw}

In the QTM{,} we calculate many independent trajectories. Because there is no relations between trajectories{,} it is easy to implement them utilizing parallel computing. Naturally, parallel computation shortens the time of calculations in comparison to serial computation. The calculated trajectories must be averaged to obtain the final trajectory.

\begin{figure}
\centering
\includegraphics[width=0.65\columnwidth]{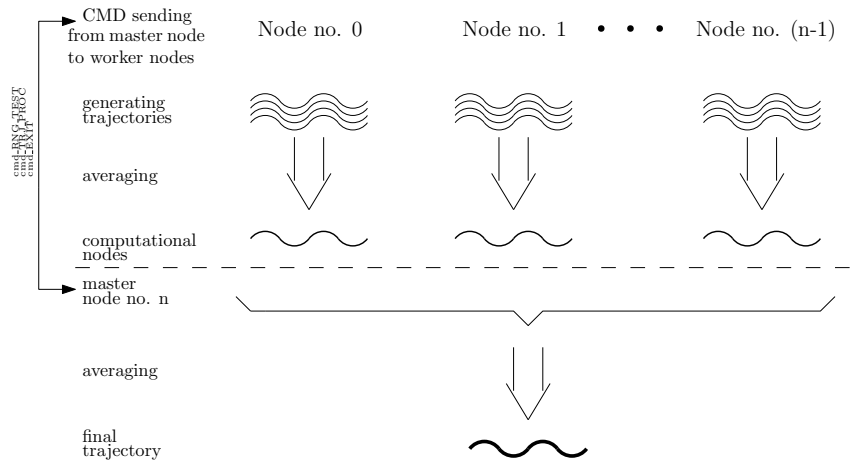}
\caption{Computing and averaging trajectories with $n+1$ computational nodes. We assume that $n$ nodes calculate trajectories{,} and one node {(denoted as node $n$)} plays a master role -- it coordinates the process of computations for the QTM. The figure presents also the general scheme of message passing during the realization of {the} QTM. The master node sends tasks, e.g. commands to calculate trajectories, and computing nodes response {with results of their calculations}}
\label{lbl:gen:and:avg:trj:cpu:PlosOne2018:ms:jw}
\label{lbl:mpi:messages:PlosOne2018:ms:jw}
\end{figure}

Fig.~\ref{lbl:gen:and:avg:trj:cpu:PlosOne2018:ms:jw} depicts the idea of computing trajectories in many computational nodes. The exchange of necessary information between nodes is realized with use of MPI protocol. This protocol offers a scalable solution{,} what means that our QTM package works efficiently within a cluster of workstations, connected with {the} use of {the Ethernet} network, and also with one multi-core personal computer. In section \ref{lbl:sec:performance:results:PlosOne2018:ms:jw}, we show the acceleration of computations carried out with {the} use of {th} QTM package -- the acceleration is noticeable regardless of whether we work with a cluster of workstations or with a one {multiprocessor} computer.

The process of calculating the final trajectory is presented {in} Fig.~\ref{lbl:mpi:messages:PlosOne2018:ms:jw}. The diagram shows the flow of messages during the calculations. The messages are sent by the master node and other nodes respond to these messages. The main part of {the} calculations is preceded by some preliminary activities, e.g. initialization of pseudorandom numbers generators. Then{,} the body of calculations begins{,} and computational nodes calculate and average trajectories. The number of trajectories is influenced by the number of computational nodes and by the number of trajectories planned to be generated to simulate an analyzed problem. We assume an uniform load for every computational node. Each computational node calculates a number of trajectories and averages them. Finally, the averaged trajectories are sent to the master node where are averaged once again {in order} to obtain the final trajectory.

Calculating {one} trajectory is directly connected with the ZVODE package. Despite the fact that this package was implemented in Fortran, the functions from {the} ZVODE library may be called in a code written in C++{,} just like other functions implemented in C++. To achieve that{,} we have to prepare an intermediary function based on a template. The exemplary function called zvode\_method\_for\_mc is presented {in} Fig.\ref{lbl:fig:zvode:code:PlosOne:ms:jw}.

\begin{figure}
\begin{lstlisting}
template <typename TYPE, size_t SIZE, size_t AlphaSize>
int zvode_method_for_mc(TYPE h, TYPE &_Tout_par, int steps,
    simpleComplex<TYPE> &_T_par,
    uVector< simpleComplex<TYPE>, SIZE > &_Y_par,
    int (*fnc)(long int *NEQ, TYPE *T, 
        dblcmplx *Y,
        dblcmplx *YDOT, 
        dblcmplx *RPAR, long int *IPAR) )
{
	int i = 0;

	int errLvl = 0;

	zvode_((fncFP)fnc,
        &neq, _Y_par.m, &_T_par.re, &_Tout_par, &itol, 
        &rtol, &atol_val, &itask, &istate, &iopt, 
        zwork, &lzw, rwork, &lrw, iwork, &liw,
        (fncFP)dummyjex,
        &mf, &rpar, &ipar);


	if(istate < 0 ) {
        if(verbose_mode == 1) 
            cerr << "HALT: Error in zvode, value ISTATE = "
                 << istate << endl ;
	    errLvl = -1;
	}

	return errLvl;
}
\end{lstlisting}
\caption{A function solving IVP (for {a} time-independent Hamiltonian) with {the} use of {the zvode\_ method} from the ZVODE package. The underscore placed in the name of {the} method is a requirement imposed by the Fortran language}
\label{lbl:fig:zvode:code:PlosOne:ms:jw}
\end{figure}

\begin{figure}
\includegraphics[width=0.95\columnwidth]{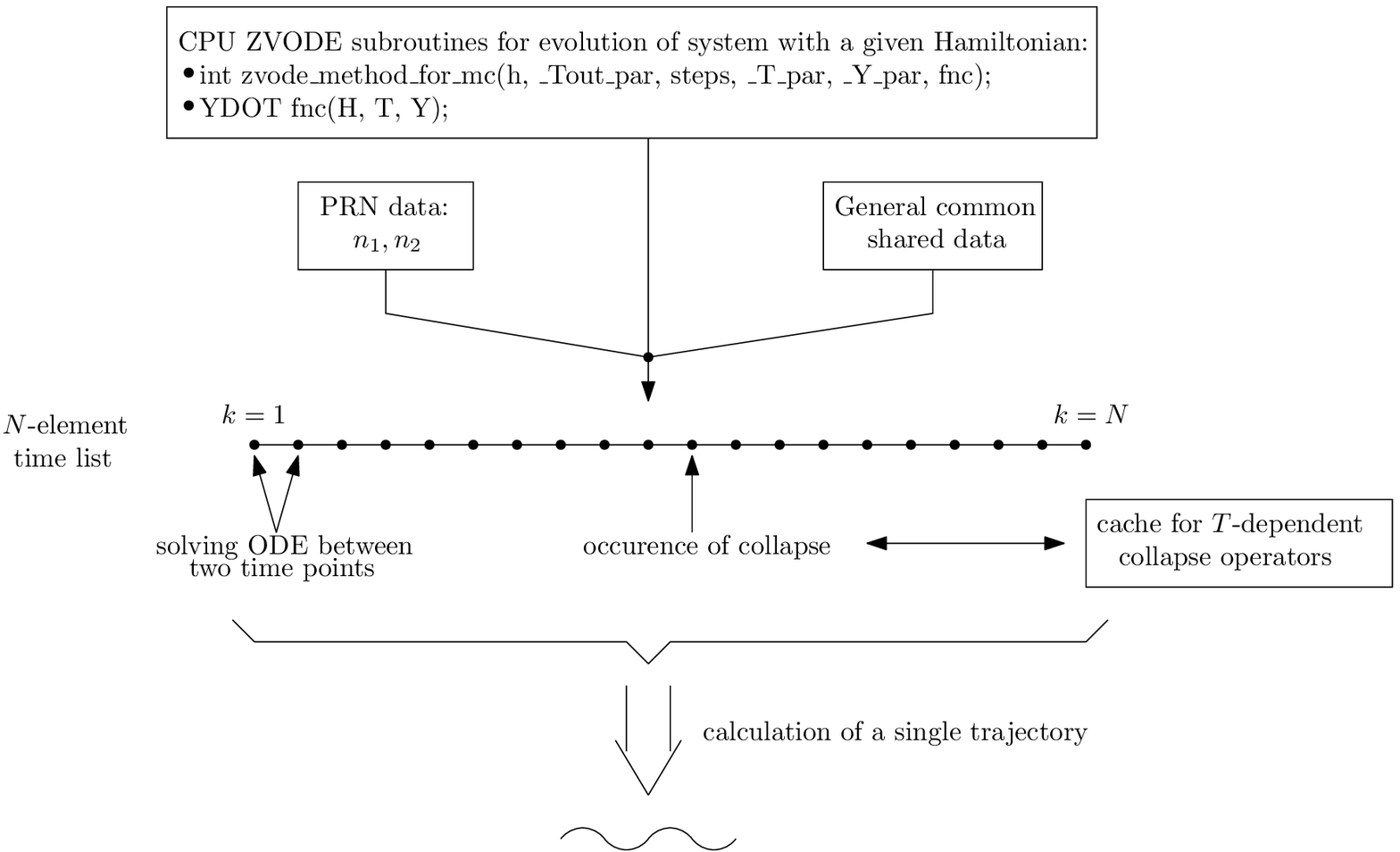}
\caption{The general scheme presenting generation of a single quantum trajectory (PRN -- pseudo-random numbers; H -- Hamiltonian data; zvode\_method\_for\_mc -- subroutine {to} ODEs solving; \_T\_par -- time variable; \_Y\_par -- actual state of the system), fnc -- calculates the right side of ODEs}
\label{lbl:gen:trj:cpu:PlosOne2018:ms:jw}
\end{figure}

The scheme of tasks realized during the calculation of a single trajectory, according to Algorithm~\ref{lbl:alg:for:single:qtm:PlosONE:ms:jw}, is shown {in} Fig.~\ref{lbl:gen:trj:cpu:PlosOne2018:ms:jw}. The algorithm is implemented in the C++ language{,} but it uses {the} zvode\_method\_for\_mc method to solve a system of ODEs. The ZVODE package does not offer the reentrant\footnote{The function's reentrant property allows many threads to utilize the same function. This property is realized by {avoidance of} using shared and global variables in function's implementation.} property, therefore{,} one MPI process may call only one instance of any method from ZVODE. However, this feature does not pose a problem because we may run many MPI processes {at} the same time. 

It is {a} very important assumption for our implementation that we use templates and static memory allocation (this applies to the code written by the user in order to, for example, describe some operators) -- these techniques {enable implementation of a} code with {an} easier declaration of objects (the code is {more} similar to a code of {a} script language than to {a} C++ code with dynamic memory allocation). This makes using the package easier{,} but equally efficient. We even expect better efficiency because the static size of data structures is known during the compilation process, therefore{,} the compiler is able to optimize the numerical computation. 

The QTM package offers three basic data types. The first one is \verb|simpleComplex<T>|. It is dedicated to the operations on complex numbers{,} where \verb|T| may be a float or a double. If we want to use \verb|simpleComplex<T>| type together with the functions from {the} ZVODE library{,} then \verb|T| is always double. The second type {\verb|uVector<T>(int size)|} is dedicated {to} vectors -- the first parameter describes the type of vector elements{,} { and a constructor parameter represents} the number of entries. Similarly for matrices{,} a type {\verb|uMatrix<T>(int size)|} was introduced{,} where the first parameter describes the type of matrix elements{,} and the constructor parameter stands for the dimension (only square matrices are utilized in QTM). { Package also utilize the static and dynamic memory allocation. However, dynamic aspects of memory management were hidden in {the} implementation layer. {Users of the} QTM package is not obliged to create objects in dynamic approach.}

{An exemplary declaration of four matrices describing collapse operators is:}
\begin{lstlisting}
uMatrix< simpleComplex<double> > c_ops[ 4 ]={ {WV_LEAD_DIM}, 
   {WV_LEAD_DIM}, {WV_LEAD_DIM}, {WV_LEAD_DIM}};
\end{lstlisting}
A value of WV\_LEAD\_DIM expresses, in this case, the dimension of operators and matrices.

In the QTM package{, a \verb|uCSRMatrix<T>(size, rowptr, colind)| type was also} implemented. This type represents a column-row oriented sparse matrix. The sparse matrices are often utilized to describe processes taking place in open quantum systems. Using sparse matrices allows {increasing} the efficiency of computation and {decreasing} the amount of memory needed to hold, for example, collapse operators or expectation values. Furthermore, sparse matrices in CSR format offer a shorter time of multiplying these matrices by vectors.

The QTM package utilizes {the} MPI standard. However, after {defining the initial structures describing a simulated problem, the} user is not obliged to deal with details of {the} MPI protocol. The whole computational process is realized by a function \verb|mpi_main|: 
\begin{lstlisting}
template<size_t N, size_t Ntrj, size_t _WV_LEAD_DIM,
         size_t _WV_LEAD_DIM_SQR, size_t _C_OPS_SIZE>
int mpi_main(int argc, char *argv[], int verbose_mode,
        double _from_time, double _to_time,
        int use_colappse_operator,
        int use_expecation_operator,
        extra_options opt);
\end{lstlisting}

In the above form of \verb|mpi_main| function{,} the Hamiltonian is time-independent. If we would like the Hamiltonian to be time-dependent, in {the} call of \verb|mpi_main|{,} we point an additional function which is used during the calculation of trajectories. The whole process of communication {with the use of MPI} is automatically realized within {the} \verb|mpi_main| function. The parameters of {the} final trajectory may be directed to standard output or to a text file{,} what is determined by a value of {the} parameter \verb|opt|. This parameter {also serves to indicate} the numerical method for solving ODEs. The ZVODE package offers two methods: ADAMS and BDF. A method is selected as below:

\begin{lstlisting}
    opt.ode_method = METBDF;
\end{lstlisting}    
or
\begin{lstlisting}
    opt.ode_method = METADAMS;
\end{lstlisting}
The BDF method is dedicated to {solving} stiff ODEs but {the ADAMS method may also} be used in the QTM (sometimes it needs to calculate more trajectories or to increase {the} accuracy of {the} Adams' ODEs solver).

\section{Selected problems -- implementation and performance} \label{lbl:sec:performance:results:PlosOne2018:ms:jw}

During the realization of {the} QTM implementation{,} the essential task was to improve its performance in comparison to other existing QTM implementations. The usage of C++ programming language and especially {the} MPI technology {facilitated us} to obtain a stable solution with decent performance{,} thanks to the parallel processing of trajectories. 

The performance of {the} presented solution was compared {with two recently developed packages {the QuTIP} and {the} QuantumOptics.jl}, which also fully supports the QTM (and {the QuTIP} also utilizes {the} ZVODE method). We have prepared two examples to examine the efficiency {of the unitary Hamiltonian and the} trilinear Hamiltonian. We also show computations referring to the Jaynes-Cummings model. The fourth example presents {the results of the experiment also conducted in \cite{Gleyzes},} and shows the accuracy of {the} QTM package. 

\subsection{Unitary Hamiltonian}

The first example concerns a simulation of a system described by the following unitary Hamiltonian:
\begin{equation}
H_{\mathrm{SYS}} = \frac{2 \pi}{10} \sigma_x, \;\;\; \mbox{and} \;\;\;  \sigma_x = \left[ \begin{array}{cc} 0 & 1 \\ 1 & 0 \end{array} \right]
\label{lbl:eqn:unitary:hamiltonian}
\end{equation}
where $\sigma_x$ represents Pauli operator $X$ (also termed as the $NOT$ operator). The initial state of {the} analyzed system is
\begin{equation}
\mket{\psi_{0}} = \mket{0} = \left[ \begin{array}{c} 1 \\ 0 \end{array} \right].
\end{equation}
The collapse operator and the {expectation} value operator, used in the simulation, are given below:
\begin{equation}
C_0 = \frac{5}{100} \sigma_x, \; E_0 = \sigma_z, \;\;\; \mathrm{and} \;\;\; \sigma_z =  \left[ \begin{array}{cc} 1 & 0 \\ 0 & -1 \end{array} \right]
\end{equation}
where $\sigma_z$ stands for {the} Pauli operator $Z$.

We ran the experiment on a PC equipped with {the} Intel i7-4950k 4.0 Ghz processor under the Ubuntu 16.04 LTS operating system. {The utilized version of the} QuTIP package is 4.2 and {the} {QuantumOptics.jl ran on Julia 0.6.4.}   
Although the dimensions of {the} above structures are quite small, the simulation of 1000 trajectories with {the QuTIP package,} when only one core is running, needs about $\approx 0.90$ seconds. Utilizing e.g. two cores for the calculation does not reduce the time of process because {the} QuTIP needs time for coordinating two threads. Of course, with the greater number of trajectories{,} it is easy to observe that the time of simulation is shorter when more cores are active.
If the QTM package is used, calculating 1000 trajectories with {the use of} e.g. of eight MPI computational nodes cores takes 2-3 seconds. This time is mainly consumed by starting the MPI processes. {It should be mentioned that {the} QuantumOptics.jl package needs more time for calculation because of {the} JIT compiler usage. However,} for a such small problems the time of calculation is almost the same, irrespective of {the} used package.

\begin{figure}
\begin{center}
\includegraphics[width=0.75\textwidth]{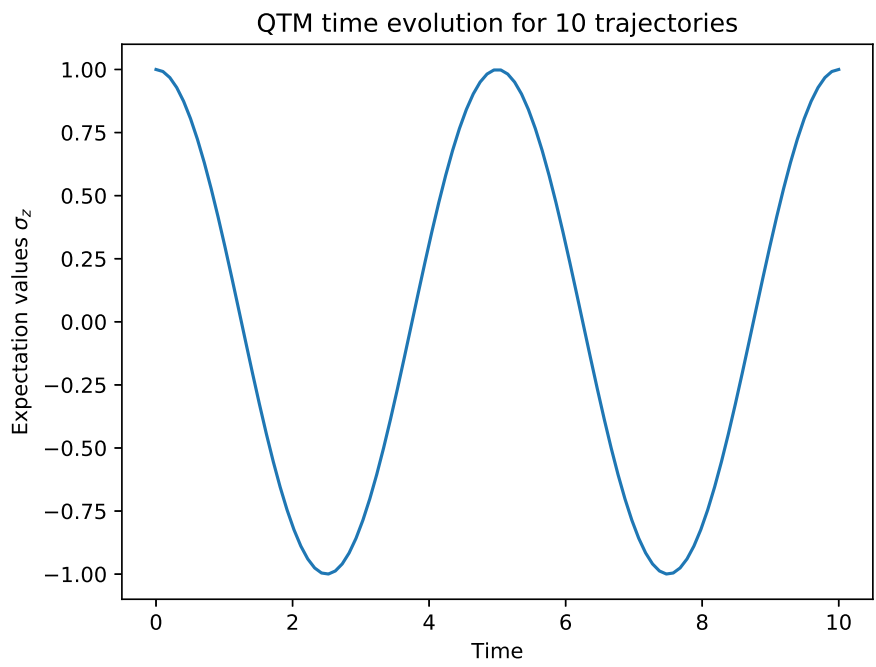}
\end{center}
\caption{The figure showing changes of {the} unitary Hamiltonian's expectation value in {the} time}
\label{lbl:fig:code:and:plot:unitary:example:PlosOne:MS:JW}
\end{figure}

Fig.~\ref{lbl:fig:code:and:plot:unitary:example:PlosOne:MS:JW} shows the evolution of {the} expectation value with {the} Pauli Z gate in time. 

The most important pieces of {the} code are written in a direct form (we define an effective Hamiltonian form directly filling {the matrix $H$ with suitable entries)} because the QTM package {allows specifying} data structures directly. Constants WD\_LD and WD\_LD\_SQR respectively stand for the vector state dimension{,} and its square and {they} are equal to two and four; co represents {the} collapse {operator}, eo stands for {the} expectation operator value and H is the Hamiltonian:
\begin{lstlisting}
int r = 0;
co[0] = msc( 0.0, 0.0 );  co[1] = msc( 0.05, 0.0 );
co[2] = msc( 0.05, 0.0 ); co[3] = msc( 0.0, 0.0 );

eo[0] = msc( 1.0, 0.0); eo[1] = msc( 0.0, 0.0);
eo[2] = msc( 0.0, 0.0); eo[3] = msc(-1.0, 0.0);

alpha[0] = msc( 1.0, 0.0);
alpha[1] = msc( 0.0, 0.0);
	
H[0] = msc( -0.00125, 0.0); 
H[1] = msc( 0.0, -0.62831853);
H[2] = msc( 0.0, -0.62831853); 
H[3] = msc( -0.00125, 0.0);
	
c_ops[0].rows=2; c_ops[0].cols=2;
c_ops[0].m = co; 

opt.type_output = OUTPUT_FILE;
opt.only_final_trj = 1;
opt.ode_method = METADAMS;
opt.tolerance = 1e-7;
opt.file_name = strdup("output-data.txt");
opt.fnc = &myfex_fnc_f1;
	
r = mpi_main<N, Ntrj, WV_LD, WV_LD_SQR, 1>(argc, 
      argv, 1, 0, 10, 1, 1, opt);
\end{lstlisting}

A significant part of the source code is a function calculating {the} right side of {the} ODEs. In the case of {the} example for a unitary Hamiltonian{, we may utilize a direct approach that is multiplying the matrix form of the Hamiltonian H by the} vector Y. The product is assigned to {the} variable YDOT. We can describe these calculation in {details} as follows:

\begin{lstlisting}
int myfex_fnc_f1(	long int *NEQ,
			double *T,
			dblcmplx *Y,
			dblcmplx *YDOT,
			dblcmplx *RPAR,
			long int *IPAR)
{
    simpleComplex<double> o0, o1, out0, out1;

      o0.re=0.0;   o0.im=0.0; o1.re=0.0;   o1.im=0.0;
    out0.re=0.0; out0.im=0; out1.re=0.0; out1.im=0.0;

    o0 = Y[0] * H[0]; o1 = Y[1] * H[1];

    out0 = o0 + o1;

    o0.re=0.0;   o0.im=0.0; o1.re=0.0;   o1.im=0.0;
    o0 = Y[0] * H[2]; o1 = Y[1] * H[3];

    out1 = o0 + o1;
    YDOT[0] = out0; YDOT[1] = out1;

    return 0;
}
\end{lstlisting}

The approach {presented above naturally allows further optimization of the code during the operation of multiplying the matrix by the} vector.

\subsection{Trilinear Hamiltonian}

In this example{, we first} assume that {the} Hamiltonian H is also time-independent.
The simulated process is a time evolution of {an} optical parametric amplifier given by the following trilinear Hamiltonian \cite{DellAnno2006}:
\begin{equation}
H_{\mathrm{SYS}} = \mathbf{i} K ( a b^{\dagger} c^{\dagger} - a^{\dagger} b c).
 \label{lbl:eqn:trilinear:hamiltonian}
\end{equation}

The symbols $a$, $b$ and $c$ stand for {the} boson annihilation operators corresponding to the pump, signal and idler fields respectively. The variable $K$ represents the value of {the} coupling constant. For the purpose of {the tests,} we assumed that $K=1${,} and $\mathbf{i}$ {represents} {an} {imaginary unit}. The initial state of analyzed system is:
\begin{equation}
\mket{\psi_{0}} = \mket{\alpha}_{a} \mket{0}_{b} \mket{0}_{c},
\end{equation}
it is a coherent state for the pump mode ($a$){, and the} vacuum states for {the signal ($b$) and the} idler ($c$) modes.

{ We also utilize three expectation operators defined as:}
{ 
\begin{gather}
a_0 = a \otimes I_b \otimes I_c, \notag \\ 
a_1 = I_a  \otimes b \otimes I_c, \notag \\
a_2 = I_a \otimes I_b \times c, \\ 
n_0 = a^{\dagger}_0 a_0, \; n_1 = a^{\dagger}_1 a_1, \; n_2 = a^{\dagger}_2 a_2 \notag  .
\label{lbl:eq:trilinear:mode:exp:op:PlosOne:2018:MS:JW}
\end{gather}
{
where $a$, $b$, $c$ still represent {the} boson annihilation operators with the same or different dimensionality. $I_a$, $I_b$ and $I_c$ are {the} identity operators for fields: pump\_mode (a), vacuum (b) and idler (c). The collapse operators are denoted as:}
\begin{gather}
c_0 = \sqrt{2 \cdot \gamma_0} a_0 \notag \\
c_1 = \sqrt{2 \cdot \gamma_1} a_1  \\
c_2 = \sqrt{2 \cdot \gamma_2} a_2 \notag \\ 
\label{lbl:eq:trilinear:mode:col:op:PlosOne:2018:MS:JW}
\end{gather}
where $\gamma_0 = \gamma_1 = 0.1, \gamma_2 = 0.4$.
}

\begin{figure}
\includegraphics[width=0.85\textwidth]{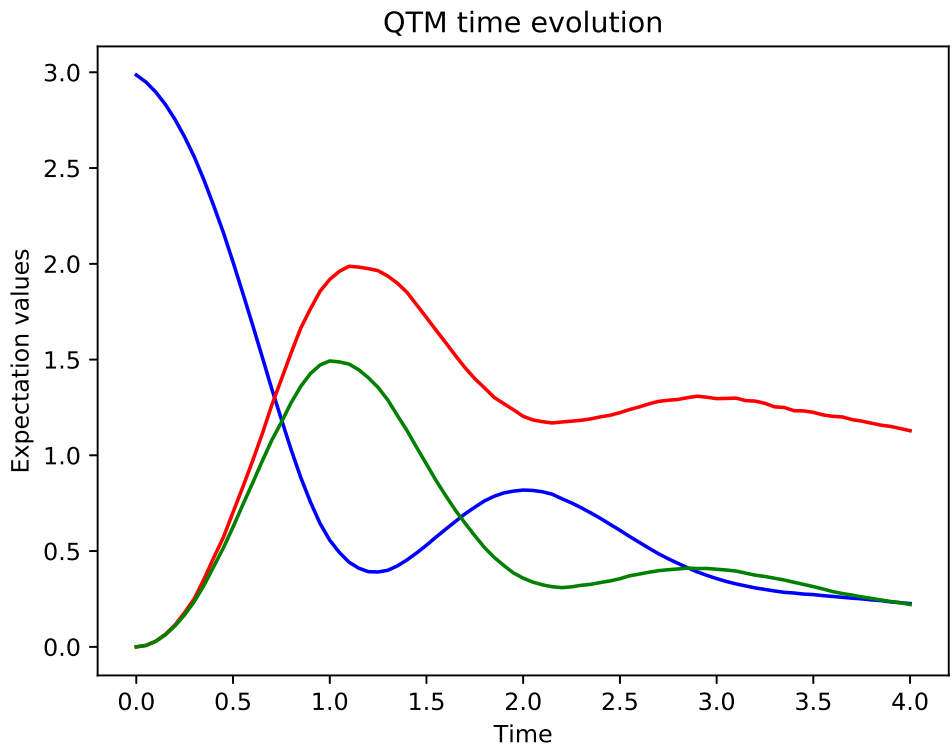}
\caption{The expectation values describing the number of photons for an experiment with generation of 1000 trajectories in three different expectation operators {(n0, n1, n2) defined {in Eq~(\ref{lbl:eq:trilinear:mode:exp:op:PlosOne:2018:MS:JW})} } for a trilinear Hamiltonian}
\label{lbl:fig:triham:photon:number:PlosOne:2018:MS:JW}
\end{figure}

The obtained expectation values of {the} photons' number, during the experiment with 1000 trajectories, are presented {in} Fig.~\ref{lbl:fig:triham:photon:number:PlosOne:2018:MS:JW}.

\begin{figure}
\includegraphics[width=0.95\textwidth]{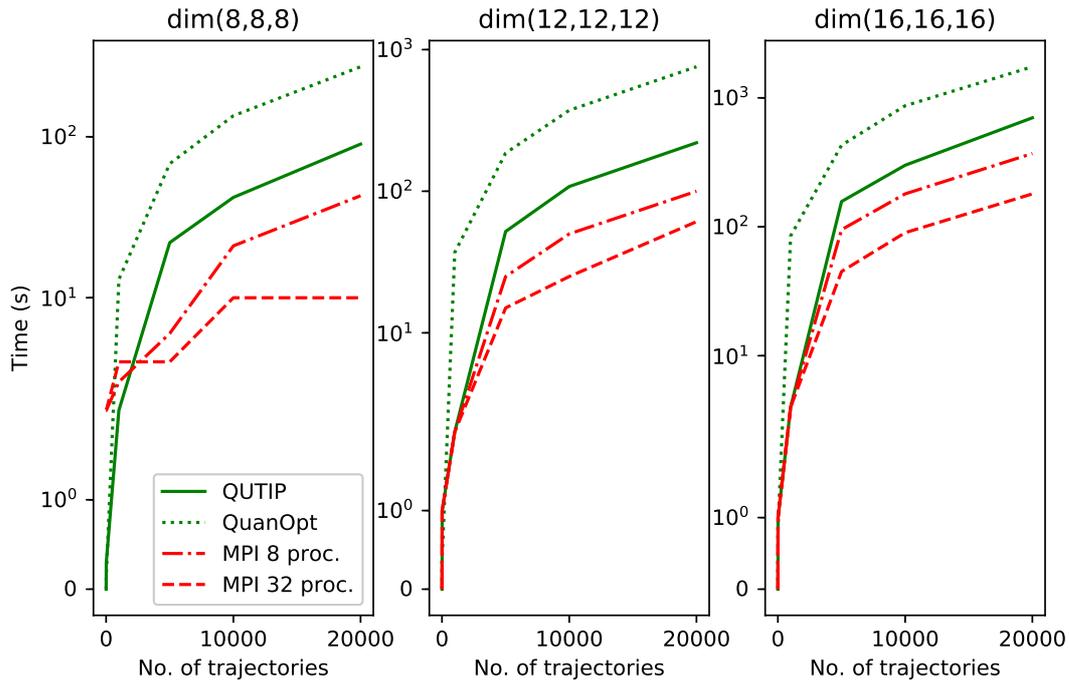}
\caption{The duration of {the} simulation for the trilinear Hamiltonian. The experiments were run for {the} different number of trajectories with use of {the} {QuTIP ( four CPU cores were used), {the} QuantumOptics.jl (termed as QuanOpt, {with only one CPU core used during the} computations){, and {the} QTM package (with two different numbers of the computational nodes)}}}
\label{lbl:fig:trilinear:hamiltonian:measure:timing:PlosOne:2018:MS:JW}
\end{figure}

Fig.~(\ref{lbl:fig:trilinear:hamiltonian:measure:timing:PlosOne:2018:MS:JW}) depicts the simulations' duration for {the} {QuTIP, {the} QuantumOptics.jl and {the} QTM packages} with {the} different number of trajectories. We also diversify the number of dimensions for the initial state in our experiments. The calculation for {the} MPI protocol was conducted with {the} use of nine computers equipped with Intel Core i7-4790K 4.0 GHZ processing units{,} and working under {the} Ubuntu 16.04 LTS operating system. Each processor has four cores, therefore{,} we were able to run {the} 32 MPI processes in eight computational nodes. The last processing unit serves as a master node.

We can observe a direct profit thanks to dividing tasks between many computational nodes. We should expect {a} significant speed-up of calculations if the number of computational nodes increases, therefore{,} we consider this {to be} a very good result. Especially for high-dimensional systems utilizing parallel computing and many computational nodes{, the execution of the tasks} directly translates into better performance i.e. shorter computation time with {the} QTM package. It should be emphasized that in {the} case of calculating 10.000 -- 20.000 trajectories 32 MPI processes still offer sufficient computing power to increase {the} size of the problem or {the} number of trajectories. For the QuTIP package{,} this computation was run with {the} use of one computational node and the whole computing power was consumed.
{The same situation {can be observed in the case of the} QuantumOptics.jl package. It also utilizes {the} JIT system of compilation and offers {a} very efficient usage of {a} processing unit. However, only calculating many trajectories using {the} MPI technology causes significant reduction of {the} calculation time.}

The simulation of {a} trilinear Hamiltonian requires using sparse matrices {in order to maintain both} low memory requirements and high performance. The QTM package offers basic data types to simplify the basic transformations and realize the calculation (also the definitions of operators may be given directly, as it was done for the Hamiltonian). A~few selected steps connected with the preparation of {the} Hamiltonian representations are presented below:

\begin{lstlisting}
// Hamiltonian preparations
const int N0=8, N1=8, N2=8;
double K=1.0, gamma0=0.1, gamma1=0.1, gamma2=0.4;
double  alpha_triham=sqrt(3);
uMatrix< simpleComplex<double> > d1(N0);
...	

uMatrix< simpleComplex<double> > a0(N0*N1*N2), C0(N0*N1*N2), num0(N0*N1*N2);
...	
uMatrix< simpleComplex<double> > H(N0*N1*N2);
...
destroy_operator(d1);
eye_of_matrix(d2);
eye_of_matrix(d3);
a0=tensor(d1,d2,d3);
...	
H=unity*K*(a0*dagger(a1)*dagger(a2)-dagger(a0)*a1*a2);
Heff=prepare_effective_H(H, C0, C1, C2);
H = convertToCSRMatrix(Heff);
...
vacuum=tensor(basis(N0,0), basis(N1, 0), basis(N2,0));
D=exp_of_matrix(alpha_triham * dagger(a0) 
                         - cojugate(alpha_triham)*a0);
alpha=D*vacuum;
...
r = mpi_main<100, Ntrj, WV_LD, WV_LD_SQR, 3>(argc, 
      argv, 1, 0, 4, 1, 1, opt);
\end{lstlisting}

The changes also apply to the function calculating {the right side of the} ODEs. Luckily, only the function realizing multiplication {has} to be changed to the one supporting the CSR matrices. Therefore, the function calculating {the right side of the} ODEs is: 

\begin{lstlisting}
int myfex_fnc_f1(long int *NEQ,
		double *T,
		dblcmplx *Y,
		dblcmplx *YDOT,
		dblcmplx *RPAR,
		long int *IPAR)
{
    size_t i, j;

    for ( i=0; i < WAVEVECTOR_LEAD_DIM; i++)
    {
        YDOT[i].re = 0.0;
        YDOT[i].im = 0.0;

        for ( j=H.row_ptr[i] ; j < H.row_ptr[i+1] ; j++)
        {
            YDOT[i]= YDOT[i] + (H.values[j] * Y[H.col_ind[j]]);
        }
    }

	return 0;
}

\end{lstlisting}

Referring to the function calculating the right side of {the} ODEs, it should be emphasized that the access to the current time value (parameter $T$) is possible{, what facilitates using a} time-dependent Hamiltonian. 

The QTM package is implemented in the C++ language and the presented examples have to be compiled, therefore{,} the computation is carried out with a high efficiency.

\subsection{Jaynes-Cummings Model}

The third example refers to the Jaynes-Cummings Model (JCM). This problem may {also be} simulated in the QTM package. In this case, a single trajectory corresponds to {a} computation run with {the} use of the master equation. Let us assume that the system's dimension is $N=40$. The dimensions of {the} operators are given in subscripts.

\begin{gather}
g=1, \; \Delta = -0.1, \; \alpha=4.0, \notag \\ 
a  = d^{-}_{N} \otimes I_2, \; b = I_N \otimes \sigma^{-}, \\
H_{\mathrm{SYS}} = \Delta a^{\dagger} a + g (a^{\dagger} b + ab^{\dagger}), \notag
\end{gather}
where $I_2$ stands for the identity operator sized $2\times 2$, $\sigma^{-}$ represents {the} annihilation operator for Pauli spins. {The values} $\Delta$ and $g$ represent coupling strength between {an} atom and {a} cavity. The initial state is expressed as:
\begin{eqnarray}
\mket{\psi_0} = \mket{\chi^{\alpha}_{N}} \otimes \mket{1}
\end{eqnarray}
where we make tensor product between a coherent state $\chi^{\alpha}_{N}$ and {a} single qubit in {the} state $\mket{1}$.
 
\begin{figure}
\includegraphics[width=0.85\columnwidth]{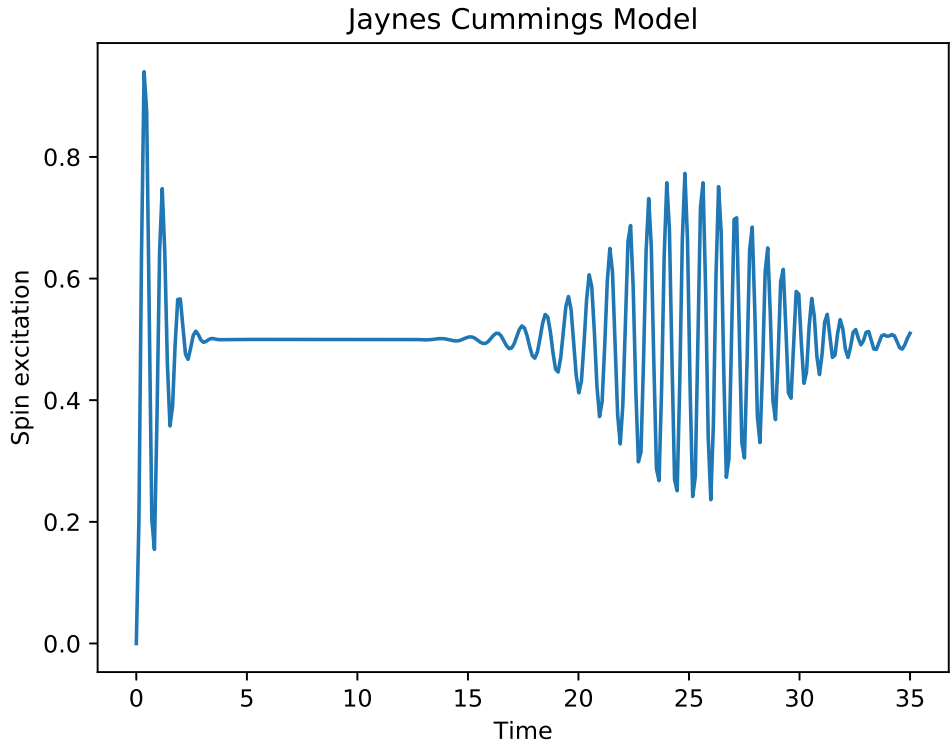}
\caption{The values of {the} spin excitation in {the} JCM obtained after calculating one trajectory without using collapse operators}
\label{lbl:fig:jcm:PlosOne:2018:MS:JW}
\end{figure}

The results obtained during the simulation of {the} JCM are presented {in} Fig.~\ref{lbl:fig:jcm:PlosOne:2018:MS:JW}. It should be emphasized that we utilize only one trajectory{,} and needed operators were represented by sparse matrices. 
 
The most important pieces of {the} code for {the} JCM are listed {as follows:}
\begin{lstlisting}
const int N = 40;
double g = 1.0, delta = -0.1, alpha = 4.0
uMatrix< simleComplex<double>, N> d1(N);
uMatrix< simleComplex<double>, 2> eye2(2), sigmam(2);
...
sigmam_operator(sigmam);
destroy_operator(d1); eye_of_matrix(eye2);
...
a = tensor(d1, eye2);
sigmaminus = tensor(eyeN, 	sigmam);
expect = dagger(sigmamnus)*sigmaminus;
H=delta * dagger(a) * a + g*(dagger(a)*sigmaminus + 
                                a * dagger(sigmaminus));
Heff=prepare_effective_H(H);
H = convertToCSRMatrix(Heff);
...
r = mpi_main<600, 1, N, N*N, 0>(argc, 
      argv, 1, 0, 35, 0, 1, opt);
\end{lstlisting}

\subsection{The birth and death of a photon in a cavity} 

The fourth and {the} last example refers to {the} convergency of the simulation of photon’s birth and death in a cavity{,} and it is based on paper \cite{Gleyzes}. Let $N$ be a number of task's dimensions{,} and $N=5$. Then{,} the Hamiltonian and {the} collapse operators {are expressed} as:

\begin{gather}
a = d^{-}, \; H_{\mathrm{SYS}}=a^{\dagger} \cdot a, \; \mket{\psi_0}= [0,1,0,0,0 ]^{\dagger}, \; \kappa = 1.0/0.129, \; \mathrm{t}=0.063, \\ \notag
c_0 = \sqrt{\kappa \cdot (1 + t)} \cdot a,  \; c_1 = \sqrt{\kappa \cdot t} \cdot a^{\dagger}
\end{gather}
where $d^{-}$ denotes the destroy operator, H -- {the} Hamiltonian and $c_0$, $c_1$ represent {the} collapse operators, t -- {the} temperature of {the} environment.

\begin{figure}
\includegraphics[width=0.95\columnwidth]{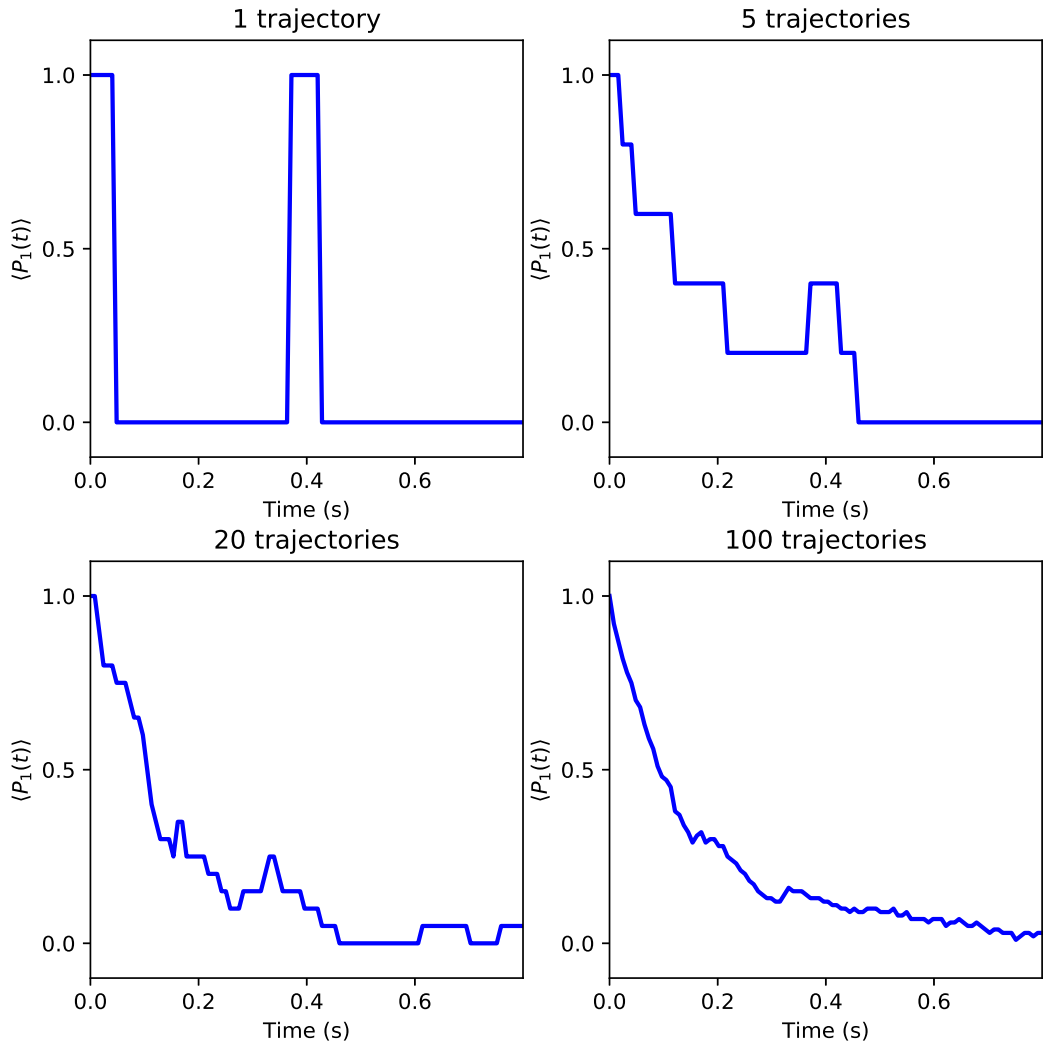}
\caption{{The} probability of {the} decay for the one-photon state obtained with {the different numbers of} trajectories. Increasing the number of trajectories improves the accuracy of calculation}
\label{lbl:fig:one:photon:trajectories:PlosOne:2018:MS:JW}
\end{figure}

Fig.~\ref{lbl:fig:one:photon:trajectories:PlosOne:2018:MS:JW} shows that the number of generated trajectories improves the accuracy for a solved problem. Naturally, it also confirms the correctness of {the} realized QTM implementation.

The source code for this example is very similar to {the} previously presented pieces of {the} code. We utilize dense matrices because the system's dimension is low{,} and it does not influence the performance.

\section{Conclusions} \label{lbl:sec:conslusions:ms:jw}

{A package which implements the Quantum Trajectories Method approach was presented in this article}. The package is dedicated to {examining the properties of the} quantum open systems. The implementation is based on {the} MPI standard. The package was prepared in the C++ programming language{,} but the implementation does not require from {the} final user any advanced programming techniques. Utilizing {the} MPI standard allows using the package within systems realizing high-performance computing{, but} also small systems like personal computers because the communication processes introduced by the MPI package {do not virtually increase demand of} computing powers.

The current version of our package is the first version that has been made public. Naturally, further works and development are planned. We would like to implement a~version based on the CUDA/OpenCL technology, which will be able to utilize computing powers of graphics processing units. 

Recently, a new approach to {the} QTM was presented \cite{Liniov2017}. In next versions of the package{,} this novelty will be considered: the version of used QTM will be matched {in accordance with a given problem in order to obtain a} shorter time of calculations.

\section*{Acknowledgments}
We would like to thank for useful discussions with {colleagues} at the Institute of Control and Computation Engineering (ISSI) of the University of Zielona G\'ora, Poland. We would like also {to thank anonymous} referees for useful comments on the preliminary version of this paper. The numerical results were done using the hardware and software available at the ''FPGA/GPU $\mu$-Lab'' located at the Institute of Control and Computation Engineering of the University of Zielona G\'ora, Poland.

\bibliographystyle{plain}

\end{document}